\documentclass[prl,twocolumn,showpacs,preprintnumbers]{revtex4} 
\usepackage{graphicx}
\usepackage{amssymb,amsmath}

\begin{document}

\title{
Exactly solvable pairing model for superconductors with a $p+ ip$-wave symmetry 
}

\author{Miguel Iba\~nez$^{(1)}$, Jon Links$^{(2)}$, 
Germ\'an Sierra$^{(1)}$, and  Shao-You Zhao$^{(2)}$}

\affiliation{$^{(1)}$Instituto de F\'{\i}sica Te\'orica, 
UAM-CSIC, Madrid, Spain} 
\affiliation{$^{(2)}$The University of Queensland, Centre for Mathematical Physics, School of Physical Sciences, 4072,
 Australia}

\bigskip\bigskip\bigskip\bigskip

%
\font\numbers=cmss12
\font\upright=cmu10 scaled\magstep1
\def\stroke{\vrule height8pt width0.4pt depth-0.1pt}
\def\topfleck{\vrule height8pt width0.5pt depth-5.9pt}
\def\botfleck{\vrule height2pt width0.5pt depth0.1pt}
\def\Zmath{\vcenter{\hbox{\numbers\rlap{\rlap{Z}\kern
0.8pt\topfleck}\kern 2.2pt
                   \rlap Z\kern 6pt\botfleck\kern 1pt}}}
\def\Qmath{\vcenter{\hbox{\upright\rlap{\rlap{Q}\kern
                   3.8pt\stroke}\phantom{Q}}}}
\def\Nmath{\vcenter{\hbox{\upright\rlap{I}\kern 1.7pt N}}}
\def\Cmath{\vcenter{\hbox{\upright\rlap{\rlap{C}\kern
                   3.8pt\stroke}\phantom{C}}}}
\def\Rmath{\vcenter{\hbox{\upright\rlap{I}\kern 1.7pt R}}}
\def\Z{\ifmmode\Zmath\else$\Zmath$\fi}
\def\Q{\ifmmode\Qmath\else$\Qmath$\fi}
\def\N{\ifmmode\Nmath\else$\Nmath$\fi}
\def\C{\ifmmode\Cmath\else$\Cmath$\fi}
\def\R{\ifmmode\Rmath\else$\Rmath$\fi}
\def\H{{\cal H}}
\def\NN{{\cal N}}
\def\tv{{\tilde{v}}}
\def\vep{{\tilde{\epsilon}}}
\def\te{{\tilde{\vep}}}
\def\sh{{\rm sh}}
\def\cth{{\rm cth}}
\def\th{{\rm th}}
\def\bk{{\bf k}}
\def\br{{\bf r}}

\begin{abstract}
We present the exact Bethe ansatz solution for the two-dimensional
 BCS pairing Hamiltonian
with $p_x + i p_y$ symmetry. Using both mean-field theory and
 the exact solution we obtain the 
ground-state phase diagram
parameterized by the filling fraction and the coupling constant.   
It consists of three phases denoted weak coupling BCS,
weak pairing, and strong pairing. The first two phases are separated
by a topologically protected line where the exact ground state is given by 
the Moore-Read pfaffian state. In the thermodynamic limit the 
ground-state energy is discontinuous on this line. The other 
two phases are separated by the
critical line, also topologically protected, previously found 
by Read and Green.  We establish 
a duality relation between
the weak and strong pairing phases, whereby ground states of the 
weak phase are ``dressed'' versions of the ground states of the strong phase 
by zero energy (Moore-Read) pairs and characterized by a topological order parameter. 
\end{abstract}

\pacs{74.20.Fg, 75.10.Jm, 71.10.Li, 73.21.La}

\preprint{IFT-UAM/CSIC-08-29}
\maketitle

\vskip 0.2cm

%
%
%
%
\def\oti{{\otimes}}
\def\lb{ \left[ }
\def\rb{ \right]  }
\def\tilde{\widetilde}
\def\bar{\overline}
\def\hat{\widehat}
\def\*{\star}
\def\[{\left[}
\def\]{\right]}
\def\({\left(}      \def\BL{\Bigr(}
\def\){\right)}     \def\BR{\Bigr)}
    \def\BBL{\lb}
    \def\BBR{\rb}
%
%
\def\zb{{\bar{z} }}
\def\zbar{{\bar{z} }}
\def\frac#1#2{{#1 \over #2}}
\def\inv#1{{1 \over #1}}
\def\half{{1 \over 2}}
\def\d{\partial}
\def\der#1{{\partial \over \partial #1}}
\def\dd#1#2{{\partial #1 \over \partial #2}}
\def\vev#1{\langle #1 \rangle}
\def\ket#1{ | #1 \rangle}
\def\rvac{\hbox{$\vert 0\rangle$}}
\def\lvac{\hbox{$\langle 0 \vert $}}
\def\2pi{\hbox{$2\pi i$}}
\def\e#1{{\rm e}^{^{\textstyle #1}}}
\def\grad#1{\,\nabla\!_{{#1}}\,}
\def\dsl{\raise.15ex\hbox{/}\kern-.57em\partial}
\def\Dsl{\,\raise.15ex\hbox{/}\mkern-.13.5mu D}
%
%
\def\ga{\gamma}     \def\Ga{\Gamma}
\def\be{\beta}
\def\al{\alpha}
\def\ep{\epsilon}
\def\vep{\varepsilon}
\def\dep{d}
\def\arc{{\rm Arctan}}
\def\la{\lambda}    \def\La{\Lambda}
\def\de{\delta}     \def\De{\Delta}
 \def\hD{{\Delta}}
\def\om{\omega}     \def\Om{\Omega}
\def\sig{\sigma}    \def\Sig{\Sigma}
\def\vphi{\varphi}
\def\sign{{\rm sign}}
\def\he{\hat{e}}
\def\hf{\hat{f}}
\def\hg{\hat{g}}
\def\ha{\hat{a}}
\def\hb{\hat{b}}
%
%
\def\CA{{\cal A}}   \def\CB{{\cal B}}   \def\CC{{\cal C}}
\def\CD{{\cal D}}   \def\CE{{\cal E}}   \def\CF{{\cal F}}
\def\CG{{\cal G}}   \def\CH{{\cal H}}   \def\CI{{\cal J}}
\def\CJ{{\cal J}}   \def\CK{{\cal K}}   \def\CL{{\cal L}}
\def\CM{{\cal M}}   \def\CN{{\cal N}}   \def\CO{{\cal O}}
\def\CP{{\cal P}}   \def\CQ{{\cal Q}}   \def\CR{{\cal R}}
\def\CS{{\cal S}}   \def\CT{{\cal T}}   \def\CU{{\cal U}}
\def\CV{{\cal V}}   \def\CW{{\cal W}}   \def\CX{{\cal X}}
\def\CY{{\cal Y}}   \def\CZ{{\cal Z}}

\def\Hp{{\mathbb{H}^2_+}} 
\def\Hm{{\mathbb{H}^2_-}}

\def\rvac{\hbox{$\vert 0\rangle$}}
\def\lvac{\hbox{$\langle 0 \vert $}}
\def\comm#1#2{ \BBL\ #1\ ,\ #2 \BBR }
\def\2pi{\hbox{$2\pi i$}}
\def\e#1{{\rm e}^{^{\textstyle #1}}}
\def\grad#1{\,\nabla\!_{{#1}}\,}
\def\dsl{\raise.15ex\hbox{/}\kern-.57em\partial}
\def\Dsl{\,\raise.15ex\hbox{/}\mkern-.13.5mu D}
%
%
%
\font\numbers=cmss12
\font\upright=cmu10 scaled\magstep1
\def\stroke{\vrule height8pt width0.4pt depth-0.1pt}
\def\topfleck{\vrule height8pt width0.5pt depth-5.9pt}
\def\botfleck{\vrule height2pt width0.5pt depth0.1pt}
\def\Zmath{\vcenter{\hbox{\numbers\rlap{\rlap{Z}\kern
0.8pt\topfleck}\kern 2.2pt
                   \rlap Z\kern 6pt\botfleck\kern 1pt}}}
\def\Qmath{\vcenter{\hbox{\upright\rlap{\rlap{Q}\kern
                   3.8pt\stroke}\phantom{Q}}}}
\def\Nmath{\vcenter{\hbox{\upright\rlap{I}\kern 1.7pt N}}}
\def\Cmath{\vcenter{\hbox{\upright\rlap{\rlap{C}\kern
                   3.8pt\stroke}\phantom{C}}}}
\def\Rmath{\vcenter{\hbox{\upright\rlap{I}\kern 1.7pt R}}}
\def\Z{\ifmmode\Zmath\else$\Zmath$\fi}
\def\Q{\ifmmode\Qmath\else$\Qmath$\fi}
\def\N{\ifmmode\Nmath\else$\Nmath$\fi}
\def\C{\ifmmode\Cmath\else$\Cmath$\fi}
\def\R{\ifmmode\Rmath\else$\Rmath$\fi}

\def\barray{\begin{eqnarray}}
\def\earray{\end{eqnarray}}
\def\beq{\begin{equation}}
\def\eeq{\end{equation}}

\def\no{\noindent}

\def\gpar{g_\parallel}
\def\gperp{g_\perp}

\def\Jb{\bar{J}}
\def\dx{\frac{d^2 x}{2\pi}}

\def\rap{\beta}
\def\s{\sigma}
\def\spec{\zeta}
\def\comb{\frac{\rap\theta}{2\pi} }
\def\Ga{\Gamma}

\def\L{{\cal L}}
\def\g{{\bf g}}
\def\K{{\cal K}}
\def\I{{\cal I}}
\def\M{{\cal M}}
\def\F{{\cal F}}

\def\gpar{g_\parallel}
\def\gperp{g_\perp}
\def\Jb{\bar{J}}
\def\dx{\frac{d^2 x}{2\pi}}
\def\imag{\Im {\it m}}
\def\real{\Re {\it e}}
\def\Jbar{{\bar{J}}}
\def\kh{{\hat{k}}}
\def\Im{{\rm Im}}
\def\Re{{\rm Re}}
\def\ti{{\tilde{\phi}}}
\def\tR{{\tilde{R}}}
\def\tS{{\tilde{S}}}
\def\tF{{\tilde{\cal F}}}
\def\ba{\bar{a}}
\def\bb{\bar{b}}
\def\be{\bar{\vep_0}}
\def\bD{\bar{\Delta_0}}

In 1957, Bardeen, Cooper and Schrieffer
published an epoch defining paper giving
a microscopic explanation of the properties of 
superconducting metals at low temperatures \cite{bcs}. The model
was based on a reduced Hamiltonian which
describes the pairing interaction between
conduction electrons. 
The original study
of the BCS model was formulated in the grand canonical
ensemble and solved with a mean-field approximation.
In 1963 
Richardson derived the exact solution of the
reduced BCS Hamiltonian with s-wave symmetry 
in the canonical ensemble \cite{richardson}. 
This solution 
was largely unnoticed until its rediscovery in the
theoretical studies of ultrasmall metallic grains 
in the 1990's, where it was employed to understand
the crossover between the fluctuation dominated
regime and the fully developed superconducting 
regime (for a review see \cite{dukelsky}).
The exact solution for the s-wave BCS model 
is related to the Gaudin spin Hamiltonians
and their integrability can be understood in the
general framework of the Quantum Inverse Scattering Method \cite{zlmg,vp}. 
These later developments allowed for an exact 
computation of various correlators \cite{zlmg,amico1,caux},
and led to  
generalizations of the Richardson-Gaudin models
with applications to condensed matter and nuclear physics 
\cite{dukelsky,dukelsky1}.

In this Letter we analyze
the two-dimensional
BCS model where the symmetry of the pairing interaction 
is $p_x + i p_y$ (hereafter referred to as $p+ip$). 
The Hamiltonian of the model is 
\begin{eqnarray}
&H& = \sum_\bk \frac{\bk^2}{2 m} \;  c^\dagger_\bk c_\bk \label{1} \\ 
&&\quad- \frac{G}{4 m} \sum_{\bk \neq \bk'}  (k_x - i k_y) (k'_x + i k'_y) \; 
 c^\dagger_\bk  c^\dagger_ {-\bk }  c_{- \bk'} c_{ \bk'}
\nonumber 
\end{eqnarray}
where $c_\bk, c^\dagger_\bk$ are destruction and creation operators
of  2D  spinless or polarised fermions with momentum $\bk$, 
$m$ is their mass and $G$ is a dimensionless coupling constant
which is positive for an attractive interaction. The $p+ip$ model has attracted considerable
attention due to the connection with the Moore-Read pfaffian state
arising in the quantum Hall effect at filling fraction
5/2 \cite{moore}, which has been proposed to support non-abelian anyons 
allowing for 
topological quantum computation \cite{tdnzz07,nssfd}. 
Motivated by these considerations, concrete proposals for engineering the $p+ip$ form of the pairing interaction have been formulated in the context of cold fermi gases \cite{ztld08,n09}. Here we will study the model through the exact Bethe ansatz solution,  
which is presented for the first time. 
We remark that exact solvability holds independent of the choice for the ultraviolet cut-off, which we denote as $\omega$, and independent of the distribution of the momenta $\bk$. In particular this means that a one-dimensional system is obtained by simply setting all $k_y=0$. Unless stated otherwise, all discussions below deal with finite particle numbers in a finite-sized system. 

Using the standard mean-field theory approach Read and Green have shown the
existence of a second-order phase transition governed
by the chemical potential $\mu$ \cite{read}. 
Adopting the terminology of \cite{read}, this transition takes
place between
a weak pairing phase ($\mu > 0$), the ground state (GS) of which behaves
as the Moore-Read pfaffian state at long distances, 
and a strong pairing phase (for $\mu < 0$).
The spectrum of Bogolioubov quasiparticles
is gapless at $\mu=0$. 
The GS of the weak pairing phase also has a non-trivial
topological structure  in ${\bf k}$-space, as shown by Volovik \cite{volovik}. 
However in the mean-field analysis the weak pairing GS is continuously connected 
to the weak coupling BCS GS \cite{read}.


Our goal is to re-examine the properties of the $p+ip$ model. Through this study we will achieve the following: i) From the mean-field results the ground-state phase diagram will be determined, comprising of the weak coupling BCS, weak pairing, and strong pairing phases; ii) A duality relation between the weak pairing and strong pairing phases will be shown to exist; iii) From the Bethe ansatz solution the duality will be formulated in terms of a dressing relation involving zero energy Moore-Read (MR) pairs; iv) Dressing of the vacuum will be seen to give the boundary line between weak coupling BCS and weak pairing phases, representing a zeroth-order quantum phase transition when the thermodynamic limit is taken (cf. \cite{maslov} for analogous zeroth-order thermal phase transitions); iv) The weak pairing phase will be shown to have a non-trivial topological structure, related to the dressing operation, which will be quantified by a winding number.       


Before presenting the exact solution of the
Hamiltonian we first extend the mean-field
results reported in \cite{read}. The  BCS order parameter associated to
(\ref{1}) is 
\beq
\hat{\Delta} = \frac{G}{m} \sum_\bk (k_x + i k_y) 
\langle  c_{- \bk} c_{ \bk} \rangle
\label{2}
\eeq
in terms of which the Hamiltonian (\ref{1}) can be approximated as (up to an additive constant)
\begin{eqnarray}
H = \sum_\bk \xi_{\bk}
\;  c^\dagger_\bk c_\bk 
- \frac{1}{4} \sum_{\bk } \left(  \hat{\Delta} \; (k_x - i k_y) 
 c^\dagger_\bk  c^\dagger_ {-\bk }   + h.c. \right) 
\label{mfham}
\end{eqnarray}
where $\xi_{\bk}= {\bk^2}/{2 m} - {\mu}/{2} $ and  $\mu/2$ is the 
chemical potential. This Hamiltonian can be
diagonalized by a Bogoliubov transformation. The gap $\Delta=|\hat{\Delta}|$
and chemical potential are the solutions of the equations 
\barray
\sum_{\bk \in {\bf K}_+ } \frac{\bk^2}{\sqrt{ (\bk^2 - \mu)^2 
+ \bk^2 \Delta^2}} & = & \frac{1}{G}
\label{4} \\
\mu \; \sum_{\bk \in {\bf K}_+  } \frac{1}{\sqrt{ (\bk^2 - \mu)^2 
+ \bk^2 \Delta^2}} & = 
& 2 M - L + \frac{1}{G} \label{4a} 
\earray
where we have set $m=1$, $L$ is the total number of energy levels and $M$ is the number 
of Cooper pairs. The set ${\bf K}_+$ denotes the 
set of momenta where $k_x > 0$ and any $k_y$, 
so that we avoid overcounting of energy levels. The mean-field expression for the GS energy is (accounting for the constant term missing in (\ref{mfham}))
\begin{eqnarray}
E=\frac{1}{2}\sum_{\bk \in {\bf K}_+} \bk^2\left(1-\frac{2\bk^2+\Delta^2-2\mu}{2\sqrt{(\bk^2 - \mu)^2 + \bk^2 \Delta^2}}\right).    
\label{gs_energy}
\end{eqnarray} 
Projection of the grand-canonical GS wave function onto a fixed number of $M$ pairs gives  
%
%
%
%
\beq
|\psi \rangle = [ \sum_{\bk \in {\bf K}_+}  g(\bk)  
c^\dagger_ {\bk } c^\dagger_ {-\bk } 
]^M 
|0 \rangle 
\label{7}
\eeq
where
%
$g(\bk) =  ({2E(\bk) - \bk^2 + \mu})/({ (k_x + i k_y) 
\hat{\Delta}^*}) 
$%
\label{8}
%
and $E(\bk)$ is the quasiparticle energy spectrum 
\beq
E(\bk) = \frac{1}{2}\sqrt{ (\bk^2 - \mu)^2 + \bk^2 \Delta^2 }. 
\label{6}
\eeq
Note that the spectrum is gapless at $\mu=0$ as $|\bk|\rightarrow 0$. 
Furthermore, the behaviour of $g(\bk)$ as 
$\bk \rightarrow 0$ depends on the sign of $\mu$ \cite{read},
\beq
g(\bk) \sim \left\{
\begin{array}{cc}
k_x - i k_y, & \mu < 0, \\
1/(k_x + i k_y), & \mu > 0. \\
\end{array}
\right.
\label{9} 
\eeq
In real space the state (\ref{7}) takes the form of a pfaffian 
\beq
\psi(\br_1, \dots, \br_{2 M}) =
{\cal A}[ g(\br_1- \br_2) \dots  g(\br_{2M-1}- \br_{2M})]
\label{10}
\eeq
where ${\cal A}$ denotes the antisymmetrization of the 
positions and $g(\br)$ is the Fourier transform of $g(\bk)$.
We will refer to the case $\mu=0$ as the Read-Green (RG) state. 
For $\mu > 0$ the large distance behaviour is $g(\br) \sim 1/(x + i y)$, 
which asymptotically reproduces the MR state 
\cite{read}.  

\begin{figure}[t!]
\begin{center}
\includegraphics[height= 4.2 cm,angle= 0]{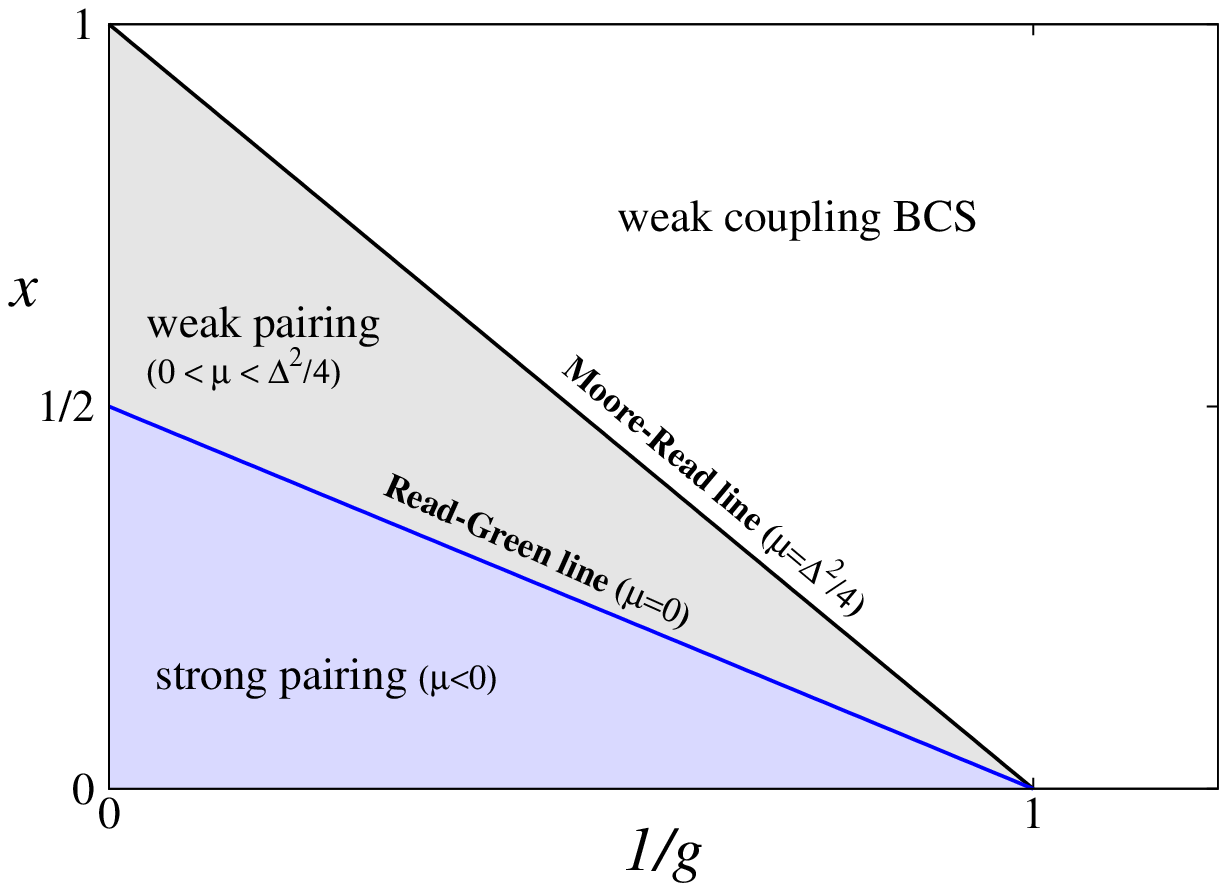}
\end{center}
\caption{
Ground-state phase diagram of the $p + i p$ 
model in terms of the inverse coupling 
$1/g$ and filling fraction $x=M/L$. Phase boundaries 
are given by the Read-Green line ($\mu=0$) and the 
Moore-Read line ($\mu=\Delta^2/4$) \cite{comment}. The phase boundaries are independent of the choice of the momentum distribution, and independent of the ultraviolet cut-off.}
\label{phase-diagram}
\end{figure}

The solution of the equations (\ref{4},\ref{4a}) for $L$ energy levels and $M$ number of pairs can be classified,
with the corresponding phase diagram  
given in fig. \ref{phase-diagram} parameterized by the filling 
fraction $x = M/L$ and the rescaled coupling constant $g = G L$. We now demonstrate how the topological aspects of the phase diagram can be deduced in a transparent manner. From (\ref{4a}) we see that $\mu=0$ imposes the relation $x_{RG}=(1-g^{-1})/2$. This result is completely independent of the momentum distribution and choice of cut-off, reflecting the topological nature of the transition discussed in \cite{read}, i.e., the boundary line is protected from perturbations  of the system  which alter the distribution of the momenta. Furthermore we identify a second topological boundary by setting $\mu=\Delta^2/4$, which from (\ref{gs_energy}) gives $E=0$, again independent of the momenta. Using (\ref{4},\ref{4a}) it is found this occurs when $x_{MR}=1-g^{-1}$. Later we will show that in this instance the GS is a discrete analogue of the MR state mentioned earlier, which in the thermodynamic limit is exactly the MR state.

A further notable $\bk$-independent property of the phase diagram
is the existence of a ``duality'' between a point
$(g, x_I)$ in the  weak pairing regime and 
another point $(g, x_{II})$ in  the strong pairing regime related
by 
\beq
x_I + x_{II} = x_{MR} \equiv  1 - \frac{1}{g}, 
\label{11}
\eeq
which necessarily can only hold for rational values of $g$. In the mean-field analysis this duality means that the
corresponding solutions are related by $\mu_I=-\mu_{II}$ and $\Delta^2_I-2\mu_I = \Delta^2_{II}-2\mu_{II}$, such that the GS energies satisfy $E_{I}=E_{II}$ according to (\ref{gs_energy}). 
 The RG state is self-dual, whereas the MR state is dual to the vacuum. This duality is apparent
in the exact solution where it will be shown to be related
to a dressing operation mentioned in the introduction.
%


%

The detailed derivation of the exact Bethe ansatz solution will be presented elsewhere. 
Here we simply mention that the technical aspects follow the derivation 
of the s-wave model solution through the Quantum Inverse Scattering 
Method, as described in \cite{zlmg,vp}. The only fundamental difference 
is that the $R$-matrix solution of the Yang--Baxter equation used to 
solve the $p+ip$ model is the trigonometric $XXZ$ solution, in contrast 
to the rational $XXX$ solution used for the s-wave model.
 
%
%
%
%
%
%
%

We again set $m=1$. The exact eigenstates of the Hamiltonian with $M$ fermion
pairs are given by 
\beq
|\psi \rangle = \prod_{j=1}^M C(y_j) |0\rangle, 
\,\,\, C(y) = \sum_{\bk \in {\bf K}_+} \frac{k_x - i k_y}{
\bk^2 - y} c^\dagger_{\bk} \; c^\dagger_{- \bk}
\label{16}
\eeq
where the rapidities $y_j,\,j=1,...,M$ satisfy the Bethe ansatz
equations (BAE)
\beq
\frac{q}{y_j} + \frac{1}{2} \sum_{\bk \in {\bf K}_+} 
\frac{1}{y_j - {\bf k}^2} - \sum_{l \neq j}^M \frac{1}{y_j - y_l} = 0,
\label{17}
\eeq
with $ 2 q= 1/G - L + 2 M - 1$. 
The total energy of the state (\ref{16}) is given by
\beq
E = ( 1 + G) \sum_{j=1}^M y_j 
\label{18}
\eeq

\begin{figure}[t!]
\begin{center}
\includegraphics[height= 5 cm,angle= 0]{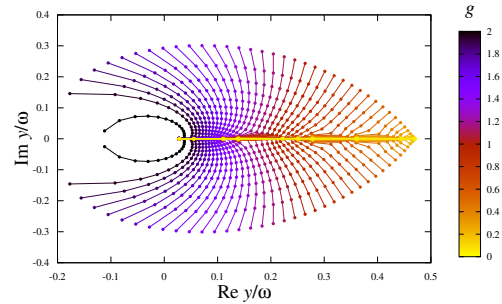}
\end{center}
\caption{
Numerical solutions for the ground-state roots $y_j \; (j=1, \dots, M)$  
of the BAEs (\ref{17})
with  $M= 31$, $L=62$, and  $0 < g < 1.99$. For this range of couplings the system is in the weak coupling BCS phase. The maximal pairing energy (the cut-off) is denoted $\omega$. 
At the critical
coupling $g=2$ (MR line) all the roots collapse to the origin (not shown).  
}
\label{arcs31}
\end{figure}

Numerical solutions of the Bethe ansatz equations indicate that there are no unpaired fermions in the GS when the fermion number is even. In fig. \ref{arcs31} we present numerical GS
solution of eqs. (\ref{17}).
This solution is obtained starting from the initial
condition $y_j \rightarrow (1 + G) \bk^2 \; (j = 1, \dots, M)$
as $G \rightarrow 0$, with the $\bk$ chosen to fill the Fermi sea. 
As $g$ increases, the roots $y_j$ closest to the
Fermi level become complex pairs.
When $g$ approaches the MR-line  the roots
bend towards the origin (as shown in fig. \ref{arcs31}) and at the value $g^{-1}=1-x$
all the roots collapse onto the origin (not shown).
At larger values of $g$ one enters the weak pairing
phase where all the roots are non-zero, except at some rational
values of $g$ where a fraction of the roots collapse
again. 
Finally, in the strong pairing regime all the roots become real
and they belong to an interval on the negative real
axis. 
%

Looking closer at the weak pairing phase,
one can check that the $M_W$ roots $y_j$ 
can be split into $M_0$ vanishing roots 
and $M_{S}$ non-zero roots provided 
%
\beq
\frac{M_0}{L} + 2 \frac{M_{S}}{L} = 1- \frac{1}{g}.
\label{24}
\eeq
Moreover the $M_{S}$  non-zero roots 
satisfy the BAE (\ref{17}) in the strong pairing 
region. Altogether this implies that given
an eigenstate, say $|S \rangle$, in the strong pairing regime
then one can dress it with $M_0$ MR pairs (as given by (\ref{24}))
obtaining an eigenstate $|W \rangle$ 
in the weak pairing phase with the same energy, 
i.e.
%
$$
H | S \rangle = E |S \rangle  \Longrightarrow
H |W \rangle = H [C(0)]^{M_0} |S \rangle = E  |W \rangle.
$$
Noticing that the filling fraction of the strong
pairing state is $x_{S} = M_{S}/L$ and that of
the weak pairing state is $x_{W} = (M_0 + M_{S})/L$,
we find that eq. (\ref{24}) coincides with 
the duality relation (\ref{11}). The physical picture
we obtain from this discussion is that the fermion
pairs forming the GS in the weak pairing phase
are of two types: strong localized pairs with negative
energy and the delocalized MR pairs with zero energy. This picture is substantially different from the projected mean-field wavefunction (\ref{7}), which is more akin to a condensate of Cooper pairs in the same one-particle state. An exception to this occurs on the MR line, where the projected mean-field and exact wavefunctions are identical. We see from (\ref{16}) that when all roots of the Bethe ansatz equations are zero, the GS is a discrete analogue of the MR state with zero energy in agreement with mean-field theory.

We reiterate that until now all our analysis has been in the context of finite-sized systems, and in particular the topological (i.e. $\bk$-independent) nature of the duality (\ref{11}) is not dependent on taking the thermodynamic limit.  In going to the thermodynamic limit we take $L,M\rightarrow \infty$, $G\rightarrow 0$ with $x=M/L$ and $g=GL$ fixed. 
A peculiar feature of the MR line is the  discontinuity 
of the GS energy $E(g,x)$ in the thermodynamic limit as the filling fraction
$x$ approaches the value $x_{MR}$ from the weak pairing region. 
To derive this result, for finite $L$
we take the one-pair state and dress it to give the dual 
GS in the weak pairing region. The filling $x_I$
of the dressed state is given by (\ref{11}), setting  $x_{II} = 1/L$, i.e.  
%
$
x_I  = x_{MR}  - {1}/{L}, 
$
%
which approaches $x_{MR} = 1 - 1/g$
as $L \rightarrow \infty$.
Since the MR pairs carry no energy, the GS 
energy of the dressed state coincides with the one-pair energy. To compute this energy
we consider the BAE for one Cooper pair
and take the continuum limit (i.e. eq. (\ref{17}) with $M=1$). Settting $\ep = {\bf k}^2$ and 
$\om$ as the cut-off, for simplicity we take the momentum distribution to be that for free particles in two dimensions, i.e. $\rho(\ep)=\omega^{-1}$. This leads to   
%
$$
L - \frac{1}{G} - 1 =  
 \sum_{\bk \in {\bf K}_+} 
\frac{y}{y - {\bf k}^2}\; 
\Longrightarrow 
1 - \frac{1}{g} = y \; \int_0^\om \frac{d \ep}{\om} 
\frac{ 1}{ y - \ep}. 
$$
This equation
has a unique negative energy solution 
$y <0$ satisfying
%
$$1 - \frac{1}{g} = \frac{y}{\om} \log \left(
\frac{ y}{y - \om} \right) 
$$
which we denote as $y = \CE(g)$. 
From here one derives the aforementioned discontinuity on the MR line $x_{MR}=1-g^{-1}$, 
%
$$\lim_{L \rightarrow \infty} E(g, x_I) = \CE(g) 
\neq E(g,x_{MR}) = 0,
$$
%
which may be described as a zeroth-order quantum 
phase transition. To the best of our knowledge, this is the 
first example of a zeroth-order quantum phase transition in 
a many-body system. 
 We have also numerically analyzed  
the excited states on the MR line
obtained by blocking the energy levels
which are occupied by unpaired electrons. These excitations
have a gap whose value agrees with the mean-field result,
suggesting that only the RG line is gapless, consistent
with mean-field theory predictions. 

As mentioned in the introduction, 
the mean-field solution shows that the weak pairing phase has a non-trivial topological
structure in ${\bf k}$-space \cite{volovik,read}. 
This structure can be characterized
by the winding number $w$  of the mean-field wavefunction
$g({\bf k}) =g_x({\bf k}) + i g_y({\bf k})$, and it is given by, 
\beq
w =  \frac{1}{\pi} \int_{\mathbb{R}^2}  d k_x \; d k_y 
\frac{ \partial_{k_x} g_x \partial_{k_y} g_y - 
\partial_{k_y} g_x \partial_{k_x} g_y }{
(1 + g_x^2 + g_y^2)^2}.
\label{w}
\eeq
One finds that  $w = 0$ for $\mu < 0$ 
(i.e. strong pairing phase), while $w=+1$ for $\mu >0$
(i.e. weak pairing and weak coupling BCS phases) \cite{volovik,read}. 
The existence
of an exact solution of the model calls for a generalization
of $w$ applicable to the many-body wavefunction
of the model $\psi({\bf k}_1, \dots, {\bf k}_M)$, where
${\bf k}_i \;(i=1, \dots, M)$ are the distinct  momenta of the pairs.  
This generalization consists in replacing $g({\bf k})$ 
in (\ref{w}) by $\psi({\bf k}+{\bf c}_1 , \dots,{\bf k}+ {\bf c}_M)$,
where ${\bf c}_j \neq {\bf c}_l \,\forall j,\l$ are a set
of distinct constants. 
With this definition we find that $w$ vanishes for the exact 
ground-state wavefunction except in the weak pairing region 
where it coincides with the number of MR pairs. Hence
$w$ provides a non-trivial topological order parameter for the weak pairing phase which is zero in the other two phases.

In summary, we have provided the exact Bethe ansatz solution for the 
BCS model with $p + ip$ pairing. Using this we have investigated 
the ground-state phase diagram,
whose structure is richer than previously supposed. We have found  that
the weak pairing region is dual to the strong pairing region,
the duality being encoded in a dressing transformation
between GS of the two phases by means of zero energy
MR pairs. The MR state obtained by dressing the vacuum
is the exact GS on a line in the phase diagram. The MR line separates 
the weak pairing and 
weak coupling BCS regions, and while the gap
does not vanish on it, the GS energy is discontinuous in 
the thermodynamic limit. We have also found a topological
order parameter that characterizes the weak pairing phase. 
An important future issue is to 
explore how  vortices (e.g. see \cite{nssfd}) can be incorporated into a 
similar model
to the one studied in this Letter. 

{\it Acknowledgments-} 
M.I. and G.S. are supported by the  CICYT project FIS2004-04885.  
G.S. also acknowledges ESF Science Programme 
INSTANS 2005-2010. J.L. and S.-Y.Z. are funded by the 
Australian Research Council through Discovery Grant DP0663772. We thank N. Read and G.E. Volovik for helpful comments.


\begin{thebibliography}{999}
\bibitem{comment} The Moore-Read line is distinct from the Higgs transition discussed in \cite{volovik1}, which occurs for $\mu=\Delta^2/2$.
\end{thebibliography}
\end{document}